\documentclass[journal]{IEEEtran}
\ifCLASSINFOpdf

\else

\fi

\usepackage{color}
\usepackage{amsfonts}
\usepackage{amsmath}
\usepackage{graphicx}
\usepackage{amssymb}
\usepackage{stfloats}
\usepackage{mathrsfs}
\DeclareMathOperator*{\argmax}{argmax}
\begin{document}

\title{Low-Complexity and High-Resolution DOA  Estimation for Hybrid Analog and Digital Massive MIMO Receive Array}
\author{Feng Shu,~
        Yaolu Qin,~
        Tingting Liu, Linqing Gui, Yijin Zhang, Jun Li, and Zhu Han
\thanks{Feng Shu,~Yaolu Qin,~Tingting Liu, Linqing Gui, Yijin Zhang, and Jun Li are with School of Electronic and Optical Engineering, Nanjing University of Science and Technology, Nanjing, 210094, China.}
\thanks{Feng Shu is also with the College of Computer and Information Sciences, Fujian Agriculture and Forestry University, Fuzhou 350002, China, and the College of Physics and Information, Fuzhou University, Fuzhou 350116, ~China.}
\thanks{Zhu Han is with the Electrical and Computer Engineering Department, University of Houston, Houston, TX 77004, USA. E-mail:  zhan2@uh.edu.}
}
\maketitle

\begin{abstract}
A large-scale fully-digital receive antenna array can provide very high-resolution direction of arrival (DOA) estimation, but resulting in a significantly high RF-chain circuit cost. Thus, a hybrid analog and digital (HAD) structure is preferred. Two phase alignment (PA) methods, HAD PA (HADPA) and hybrid digital and analog PA (HDAPA), are proposed to estimate DOA based on the parametric method. Compared to analog phase alignment (APA), they can significantly reduce the complexity in the PA phases. Subsequently, a fast root multiple signal classification HDAPA (Root-MUSIC-HDAPA) method is proposed specially for this hybrid structure to implement an approximately analytical solution. Due to the HAD structure, there exists the effect of direction-finding ambiguity. A smart strategy of maximizing the average receive power is adopted to delete those spurious solutions and preserve the true optimal solution by linear searching over a set of limited finite candidate directions. This results in a significant reduction in computational complexity.  Eventually, the Cramer-Rao lower bound (CRLB) of finding emitter direction using the HAD structure is derived. Simulation results show that our proposed methods, Root-MUSIC-HDAPA and HDAPA, can achieve the hybrid CRLB with their complexities being significantly lower than those of pure linear searching-based methods, such as APA.
\end{abstract}
\begin{IEEEkeywords}
DOA, Hybrid Analog and Digital, massive MIMO, CRLB, Root-MUSIC-HDAPA.
\end{IEEEkeywords}
\section{Introduction}
Due to ultra-high-resolution of spatial direction, and super-high-spectral efficiency, massive multiple-input multiple-output (MIMO) has drawn tremendous research activities from academia and industry world. It has made great progress on several important aspects like channel modeling, low-complexity beamforming, channel estimation, pilot optimization, pilot contamination controlling, etc. \cite{Rusek}, \cite{Marzetta}, \cite{Zhao}, \cite{Ma}.  Direction of arrival (DOA) estimation has been an active area since its applications include wireless communications, radar, radio astronomy, sonar, navigation, tracking of various objects, and rescue and other emergency assistance devices \cite{Tuncer}. If massive MIMO behaves as a receive array,  then DOA estimation precision will be dramatically improved due to its ultra-high-resolution of spatial direction.

Wireless direction finding has a long history tracing back to the very beginnings of wireless communications.  In the  coming future, demand for direction-finding will arise in many potential engineering applications including internet of things (IoT) \cite{Kaurn}, directional modulation systems \cite{Hu1,Shu1,Shu2,Hu2,Ding,kalantari,Nusenu}, unmanned aerial vehicle (UAV) \cite{Zengy}, intelligent transportation, wireless sensor networks (WSNs) \cite{Fei} and millimeter-wave-based massive MIMO for 5G and beyond so on \cite{yiwang}.  Many DOA estimation algorithms have been proposed and analyzed. Capon algorithm \cite{Capon} is maximum likelihood estimation of power which aims at maximizing the signal-to-interference ratio (SINR). Schmidt developed a more popular method, i.e., the multiple signal classification (MUSIC) \cite{Schmidt} algorithm, which is a high-resolution eigen-structure-based DOA-finding method. To reduce the complexity of  MUSIC with linear search, its low-complexity version, called Root-MUSIC \cite{Barabell}, was proposed to solve the roots  of the polynomial around the unit circle to find DOA. Many authors proposed several different direction finding methods,  analyzed and improved their performances. In \cite{Stoica}, five methods of combining ML and MUSIC were proposed, which could achieve both good performance and computational simplicity. In \cite{Wang}, the authors exploited the first derivative of the cost function in Root-MUSIC,  which performed better than traditional methods. By examining the disturbance of the root of the polynomial formed on the Root-MUSIC intermediate step, the authors in \cite{Rao} provided its analysis and proved that it outperforms the MUSIC algorithm in a uniformly spaced linear array (ULA). However, the above research all assume that array response and noise variance are known perfectly, which is unfeasible in practice. Therefore, Friedlander modeled and solved the problem of  direction finding  when there existed inaccurate mutual coupling, gain, and phase  among array elements \cite{Friedlander1, Scharf, Friedlander2}.

However, as the number of antennas tends to large-scale, the beamforming computational amount, and circuit complexity and cost of  digital implementation become too high for commercial applications. Therefore, a hybrid analog and digital (HAD) beamforming structure is a natural choice, which will strike a good balance among beamforming computational amount, circuit cost, and circuit implementation complexity.
Concerning HAD precoding in mmWave massive MIMO systems,  a mixed analog-to-digital converter (ADC) receiver architecture \cite{Tan} was presented, as combining costly high and less expensive low resolution ADCs, had worse in performance than full-resolution ADC structure. Therefore, in \cite{Zhang}, a HAD precoding algorithm is firstly proposed  to make a balance between hardware cost and system performance.

Several research activities on HAD structure focus on transmitter not receiver.  In \cite{Yu}, the authors developed a low-complexity  precoder of alternately iterative minimization by enforcing an orthogonal constraint on the digital precoder. In \cite{Alluhaibi}, the authors proposed two precoders based on the principle of manifold optimisation and particle swarm optimisation. An energy-efficient hybrid precoding for sub-connected architecture was proposed in \cite{Gao}. To make a balance between energy efficiency and spectrum efficiency in \cite{Han}, the authors analyzed the green point for fixed product of the transceivers number and the active antennas number per transceiver, and independent transceivers number and active antennas number per transceiver. Due to the HAD structure, the achievable sum-rate inevitably decreases compared to fully-digital beamforming, in \cite{Ying} the sum-rate degradation was proved to be compensated by simply employing more transmit antennas. Also taking the rate into account, the authors in \cite{Alkhateeb} developed an iterative HAD beamforming algorithm for the single user mmWave channel, which can approach the rates achieved by unconstrained digital beamforming solutions. In  \cite{Ayach}, the authors presented receive baseband combiners with the target of minimizing mean-squared-error between transmitted and processed received signals.

Medium-scale or large-scale receive antenna array with digital beamforming can be employed at receiver to achieve a high-resolution DOA estimation. Therefore, considering the hardware cost and performance, it is necessary to apply the hybrid structure in the direction finding. In \cite{Huang}, the authors proposed two iteration methods, i.e., differential beam search and differential beam tracking beamforming algorithms for side by side subarray configuration.

To the best of our knowledge, how to use a massive HAD beamforming structure to make an estimate of DOA direction based on concept of spatial spectrum is an open  challenging problem. In this paper,  each subarray output of the HAD structure is viewed as a virtual large antenna output,  and the total HAD antenna array can be modelled as a large digital virtual array when we do digital beamforming/PA operation.  we will focus on the aspect research and make our effort to solve this problem, our main contributions are summarized as follows:
\begin{enumerate}
  \item By fully exploiting the sub-array structure, two  hybrid DOA-finding methods of using linear search, hybrid analog and digital phase alignment (HADPA) and  hybrid  digital and analog phase alignment (HDAPA), are proposed to estimate DOA based on parametric method. Compared to conventional analog phase alignment (APA), they are much lower-complexity. By reducing the size of stepsize,  their estimate accuracy can be improved but at the same time their complexities increase accordingly. Compared to APA, the proposed HADPA can  reduce the complexity from $O(KM)$ to $O(K+M)$, where $K$  and $M$ are the numbers of subarrays  and antennas per subarray. Furthermore, the proposed HDAPA dramatically reduces the search complexity by confining the searching set of feasible solutions to the limited finite number $M$ by exploiting the the periodic characteristic of digitally large virtual array with spacing being multiple of half wavelength.

 \item To avoid the high-complexity of HADPA and APA due to pure linear searching with small stepsize, based on spatial spectral estimation, a  Root-MUSIC-HDAPA method is proposed to achieve an extremely low-complexity with an approximately close form. Due to the periodic property of virtual array direction pattern, there exists the effect of direction-finding ambiguity effect, i.e., $M$ optimal solutions for the estimated direction.  A method of maximizing the average receive power by a limited linear searching over a set of finite feasible directions predetermined by  Root-MUSIC, called HDAPA,  is adopted to find the true optimal solution and delete those spurious ones. As shown in mathematic analysis and simulation results in Section V, Root-MUSIC plus HDAPA can make a significant reduction in computational complexity compared with APA, HADPA, and HDAPA.

 \item To assess the performance of the proposed two methods,  the hybrid Cramer-Rao lower bound (CRLB) for HAD  structure is derived by statistic theory and matrix theory. Simulation results verify that  the proposed hybrid Root-MUSIC-HDAPA scheme is shown to achieve the CRLB as signal-to-noise ratio (SNR)  increases up to medium and large SNR regions.
\end{enumerate}

The remainder of this paper is organized as follows. Section \ref{S2} describes  system model. In Section \ref{S3}, two methods, HADPA and HDAPA, are proposed to realize a lower-complexity compared to conventional APA. In Section \ref{S4}, compared to HADPA and HDAPA, a lower-complexity  Root-MUSIC-HDAPA is proposed by providing an approximately analytical solution, and the corresponding hybrid CRLB is also derived to verify its performance. Simulation results are presented  in Section \ref{S5}. Finally, we make our conclusions in Section \ref{S6}.

Notation: throughout the paper, matrices, vectors, and scalars are denoted by letters of bold upper case, bold lower case, and lower case, respectively. Signs $(\cdot)^T$, $(\cdot)^*$£¬ and $(\cdot)^H$ denote transpose,~conjugate,~and conjugate transpose,~respectively. Notation $\mathbb{E}\{\cdot\}$ stands for the expectation operation. Matrices $\textbf{I}_N$ denotes the $N\times N$ identity matrix and $\textbf{0}_{M\times N}$ denotes $M\times N$ matrix of all zeros. $\mathrm{Tr}(\cdot)$ denotes matrix trace. Operation $\otimes$ denotes the Kronecker product of two matrices.

\section{System Model}\label{S2}

\begin{figure}[h]
\centering
\includegraphics[width=0.5\textwidth]{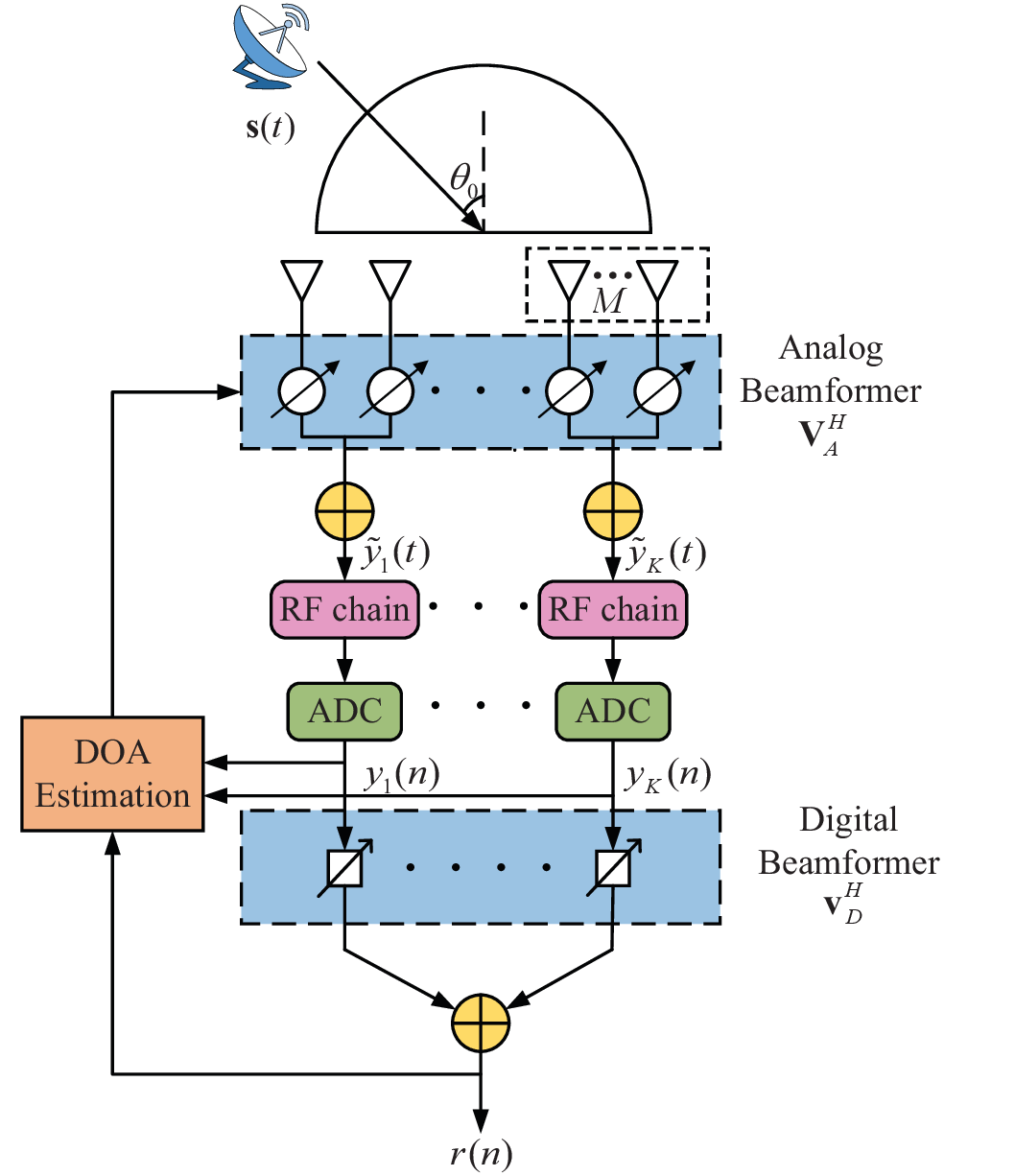}\\
\caption{ULA hybrid beamforming sub-connected architecture.}\label{Visio-jiegou}
\end{figure}

Fig.~\ref{Visio-jiegou} sketches the receive HAD beamforming structure. A far-field  emitter transmit a narrow-band signal $s(t)e^{j2\pi f_ct}$, where $s(t)$ is the baseband signal, and $ f_c$ is the carrier frequency.  The signal impinges on the HAD antenna array. Uniformly-spaced linear  array (ULA)  are divided into $K$ subarrays, and each subarray is composed of $M$ antenna elements. Consider analog beamforming (AB),  the $k$th subarray output $\tilde{y}_k^{b}(t)$ is
\begin{align}\label{apa_output¡ªk}
\tilde{y}_k^{b}(t)=\sum^{M}_{m=1}s(t)e^{j\big(2\pi f_ct-2\pi f_c\tau_{k,m}-\alpha_{k,m}\big)}+w_k^b(t), 1\le k\le K,
\end{align}
where $b$ is the time-domain block index with each block consisting of $L$ snapsots, i.e., $L$ is the number of snapshots per block, and $\tau_{k,m}$ are the propagation delays determined by the direction of the source relative to the array given by
\begin{align}
\tau_{k,m}=\tau_0-\frac{(km-1)d\sin{\theta_0}}{c},
\end{align}
where $\tau_0$ is the propagation delay from the emitter to a reference point on the array, $c$ is the speed of light, and $d$ denotes the antenna spacing. In (\ref{apa_output¡ªk}), $\alpha_{k,m}$ is the corresponding phase  for analog beamforming/phase alignment corresponding to the $m$th antenna of subarray $k$.  Stacking all $K$ subarray outputs in (\ref{apa_output¡ªk}) forms the matrix-vector notation
\begin{align}\label{apa_output¡ªvec}
\mathbf{\tilde {y}}^b\left(t\right)=e^{j2\pi f_ct}\mathbf{V}_{A}^{H}\mathbf{a}(\theta_0)s(t)+\mathbf{w}^b(t),
\end{align}
where $\mathbf{w}(t)=\left[w_{1}(t),w_{2}(t),\cdots,w_{K}(t)\right]^{T}$ is an additive white Gaussian noise (AWGN), whose entries are independent identically distributed. $\mathcal{CN}\left(0,\sigma_{w}^{2}\right)$, and the column vector $\mathbf{a}(\theta_0)$ is the so-called array manifold defined by
\begin{align}\label{d}
\mathbf{a}(\theta_0)=\left[1,e^{j\frac{2\pi}{\lambda}d\sin{\theta_0}},\cdots,e^{j\frac{2\pi}{\lambda}\left(N-1\right)d\sin{\theta_0}}\right]^{T},
\end{align}
and the AB matrix is a block diagonal matrix
\begin{align}
\mathbf{V}_{A}
=\begin{bmatrix}
\mathbf{v}_{A,1} & \mathbf{0} & \cdots & \mathbf{0}\\
\mathbf{0} & \mathbf{v}_{A,2} & \cdots & \mathbf{0}\\
\vdots & \vdots & \ddots & \vdots\\
\mathbf{0} & \mathbf{0} & \cdots & \mathbf{v}_{A,K}\\
\end{bmatrix}
\end{align}
where $\mathbf{v}_{A, k}=\frac{1}{\sqrt{M}}\left[e^{j\alpha_{k,1}},e^{j\alpha_{k,2}},\cdots,e^{j\alpha_{k,M}}\right]^{T}$ is the AB vector of subarray $k$. The radio frequency (RF) signal $\mathbf{\tilde {y}}\left(t\right)$ in (\ref{apa_output¡ªvec}) passes through $K$ parallel RF chains and is down-converted to the following baseband signal vector
\begin{align}
\mathbf{y}^{b}(t)=\mathbf{V}_{A}^{H}\mathbf{a}(\theta_0)s(t)+\mathbf{w}^b(t),
\end{align}
which experiences analog-to-digital convertor (ADC) and yields
\begin{align}
\mathbf{y}^{b}(n)=\mathbf{V}_{A}^{H}\mathbf{a}(\theta_0){s}(n)+\mathbf{w}^b(n).
\end{align}
Via digital beamforming (DB) operation, the above signal vector becomes
\begin{align}\label{r}
r^{b}(n)=\mathbf{v}_{D}^{H}\mathbf{V}_{A}^{H}\mathbf{a}(\theta_0){s}(n)+\mathbf{v}_{D}^{H}\mathbf{w}^b(n),
\end{align}
where the DB vector $\mathbf{v}_{D}=\left[v_1, v_2, \cdots, v_K\right]^{T}$.

\section{Proposed Low-Complexity Phase-Alignment-based DOA Estimation}\label{S3}
In this section, by maximizing the output receive power, we firstly present the APA method in Section \ref{S_APA}. The APA requires to compute $N$ different values of phase per step search as the inputs of $N$ phase shifters at the RF chain. To reduce the number of phase values needed to be computed, a low-complexity HADPA-based DOA estimator is proposed in Section \ref{S_HADPA} with only calculating $M+K$ different values for $N$ phase shifters on RF chain. This significantly alleviates the computational load at receive terminal. In order to further reduce  complexity, we finally propose a HDAPA DOA estimator in Section \ref{S_HDAPA}, making each-step search only requires $K$ values of phases. By exploiting  the periodic characteristics of the large virtual digital array, we obtain a feasible set of $M$ estimated  angles, and then  the APA is used to delete the false angles and keep the true optimal angle.
\subsection{Conventional APA}\label{S_APA}
It is assumed that the emitter direction in Fig.~1 is $\theta_0$. From the previous section,  after AB and ADC, we have  the output summation signal of the $k$th subarray as follows
\begin{align}
y_{k}^{b}(n)&=\mathbf{v}_{A,k}^H\mathbf{a}_k(\theta_0)s(n)+w_k^b(n)\\\nonumber
&=\frac{1}{\sqrt{M}}s(n)e^{j\frac{2\pi}{\lambda}(k-1)Md\sin{\theta_0}}\times\\\nonumber
&\sum_{i=1}^{M}e^{j\left(\frac{2\pi}{\lambda}(m-1)d\sin\theta_0-\alpha_{k,m}\right)}+w_k^b(n),
\end{align}
where $\mathbf{a}_k(\theta_0)$ is the array manifold of subarray $k$,
\begin{align}
\mathbf{a}_k(\theta_0)=\left[
e^{j\frac{2\pi}{\lambda}(k-1)Md\sin{\theta_0}},\cdots,e^{j\frac{2\pi}{\lambda}(kM-1)d\sin{\theta_0}}
\right]^T.
\end{align}
 Since only APA is used, the DB vector $\mathbf{v}_{D}$ is set and fixed to  a vector of all ones, i.e., $\mathbf{v}_{D}=[1,1,\cdots,1]^{T}$.

Below, we will maximize the output power of receive signal $r(n)$ in Fig.~\ref{Visio-jiegou} by optimizing the vector $\mathbf{v}_{A,k}$.  Firstly, let us define the average output power
\begin{align}
P^b=\frac{1}{L}\sum^{L}_{n=1}[r^b(n)r^b(n)^{H}]=\frac{1}{L}\mathbf{r}^b{\mathbf{r}^b}^H,
\end{align}
where $\mathbf{r}^{b}=\left[r^b(1),\cdots,r^b(L)\right]$. The above equation can further be expanded as
\begin{align}\label{P_r_APA}
P_r^b(\theta)&=\frac{1}{LN^2}\sum_{n=1}^{L}\left[r^b(n)r^b(n)^{H}\right]\\\nonumber
&=\frac{1}{LN^2}\sum_{n=1}^{L}\left[(\sum_{k=1}^{K}y_k^b(n))(\sum_{k=1}^{K}y_k^b(n))^H\right]\\\nonumber
&=\frac{1}{LN^2}\sum_{n=1}^{L}[(\sum_{k=1}^{K}\frac{1}{\sqrt{M}}s(n)e^{j\frac{2\pi}{\lambda}(k-1)Md\sin{\theta_0}}\times\\\nonumber
&\sum_{i=1}^{M}e^{j\left(\frac{2\pi}{\lambda}(m-1)d\sin{\theta_0}-\alpha_{k,m}\right)}+w_k^b(n))\times\\\nonumber
&(\sum_{k=1}^{K}\frac{1}{\sqrt{M}}s(n)e^{j\frac{2\pi}{\lambda}(k-1)Md\sin{\theta_0}}\times\\\nonumber
&\sum_{i=1}^{M}e^{j\left(\frac{2\pi}{\lambda}(m-1)d\sin{\theta_0}-\alpha_{k,m}\right)}+w_k^b(n))^{H}],
\end{align}
where
\begin{align}\label{alpha}
\alpha_{k,m}=\frac{2\pi}{\lambda}\left((k-1)M+(m-1)\right)d\sin{\theta}.
\end{align}

By adjusting the value of $\theta$ in (\ref{alpha}), we can optimize the receive power in (\ref{P_r_APA}) to reach its maximum value. Observing the last line of (\ref{P_r_APA}), we find  the analog optimizing vector $\mathbf{v}_{A,k}$ is exactly aligned with the array manifold produced by the direction $\theta_0$  under the condition
\begin{align}\label{alpha_opt}
\alpha_{k,m}=\frac{2\pi}{\lambda}\left((k-1)M+(m-1)\right)d\sin{\theta_0},
\end{align}
which forces all signals of $N$ antenna elements to coherently combine at the output and form the maximum value of output power. To implement linear exhaustive searching, we split the range of direction angle $\theta$  from $-\frac{\pi}{2}$ to $\frac{\pi}{2}$ into $Q$ subintervals or bins. Let us define the phase searching stepsize as follows
\begin{align}
\Delta\theta=\frac{\pi}{Q}.
\end{align}
In (\ref{alpha}), the angle $\theta$ is chosen from the angle set $\Theta=\{-\pi/2,-\pi/2+\Delta\theta,\cdots,\pi/2\}$. As the search direction angle $\theta$ varies from $-\pi/2$ to $\pi/2$, the APA before ADC at the receiver in Fig.~1 cannot save the receive data unlike DPA. In other words, the new APA phases should be computed and sent towards $N$ phase shifters per step-search and the new block of signal will be received to compute the output outcome of the new search point. The total number of values of all $P_r^b(\theta)'s$ are $L(Q+1)KM$ floating-point operations (FLOPs). Thus the complexity of APA is
\begin{align}\label{complexAPA}
C_{APA}=O(L(Q+1)KM)
\end{align}
FLOPs. Finally, the maximum receive power are found by comparison. Obviously, to approach the CRLB,  the stepsize $\Delta\theta$ should be chosen such small that it is close to the root of CRLB. This implies a large value of $Q$ and a high computational amount.

\subsection{Proposed Low-Complexity HADPA DOA Estimator}\label{S_HADPA}
The above APA algorithm needs to do exhaustive linear search from $-\frac{\pi}{2}$ to $\frac{\pi}{2}$ and compute $N$ values at the same time, which will cause a high-complexity. In the subsection, we present a low-complex hybrid phase alignment DOA estimation. Firstly, we decompose the PA phase $\alpha_{k,m}$ into two parts:
\begin{align}
\alpha_{k,m}=\alpha_{m}+\alpha_{k},
\end{align}
where the first part is to cancel the phase of  element $m$ for each subarray and the second part $\alpha_{k}$ is to cancel the common phase of subarray $k$. This means PA consists of two steps: APA in the first step and DPA in the second step.

After the APA, the output of subarray $k$ is described as follows
\begin{align}
y_{k}^b(n)&=\mathbf{v}_{A,k}^H\mathbf{a}_k(\theta_0)s(n)+w_k^b(n)\\\nonumber
&=\frac{1}{\sqrt{M}}s(n)\underbrace{e^{j\frac{2\pi}{\lambda}(k-1)Md\sin{\theta_0}}}_{\text{Common~factor~for subarray}~k}\times\\\nonumber
&\sum_{m=1}^{M}e^{j\left(\frac{2\pi}{\lambda}(m-1)d\sin\theta_0-\alpha_{m}\right)}+w^b_k(n),
\end{align}
where
\begin{align}\label{a_m}
\alpha_{m}=\frac{2\pi}{\lambda}(m-1)d\sin{\theta}.
\end{align}
To remove the common factor of subarray $k$, we design the following DPA  vector
\begin{align}
\mathbf{v}_D=[e^{j\alpha_{1}},e^{j\alpha_{2}},\cdots,e^{j\alpha_{K}}]^H,
\end{align}
where
\begin{align}\label{a_k}
\alpha_{k}=\frac{2\pi}{\lambda}(k-1)Md\sin{\theta}.
\end{align}
Therefore, $r^b(n)$ in Fig.~1 is represented as
\begin{align}\label{r2}
r^b(n)&=\sum_{k=1}^{K}e^{-j\alpha_{k}}y_{k}^b(n)\\\nonumber
&=\frac{1}{\sqrt{M}}s(n)\sum_{k=1}^{K}e^{j(\frac{2\pi}{\lambda}(k-1)Md\sin{\theta_0}-\alpha_{k})}\\\nonumber
&\times{\sum_{m=1}^{M}e^{j(\frac{2\pi}{\lambda}(m-1)d\sin{\theta_0}-\alpha_{m})}}+\sum_{k=1}^{K}e^{-j\alpha_k}w_k^b(n).
\end{align}
Similar to (\ref{P_r_APA}), we have the average receive power as follows
\begin{align}\label{P_r_HADPA}
P_r^b(\theta)&=\frac{1}{LN^2}\sum_{n=1}^{L}[r^b(n)r^b(n)^{H}]\\\nonumber
&=\frac{1}{LN^2}\sum_{n=1}^{L}[\big(\frac{1}{\sqrt{M}}s(n)\sum_{k=1}^{K}e^{j(\frac{2\pi}{\lambda}(k-1)Md\sin{\theta_0}-\alpha_{k})}\\\nonumber
&\times{\sum_{m=1}^{M}e^{j(\frac{2\pi}{\lambda}(m-1)d\sin{\theta_0}-\alpha_{m})}}+\sum_{k=1}^{K}e^{-j\alpha_k}w_k^b(n)\big)\\\nonumber
&\times\big(\frac{1}{\sqrt{M}}s(n)\sum_{k=1}^{K}e^{j(\frac{2\pi}{\lambda}(k-1)Md\sin{\theta_0}-\alpha_{k})}\\\nonumber
&\times{\sum_{m=1}^{M}e^{j(\frac{2\pi}{\lambda}(m-1)d\sin{\theta_0}-\alpha_{m})}}+\sum_{k=1}^{K}e^{-j\alpha_k}w_k^b(n)\big)^{H}].
\end{align}
According to the APA mentioned above, we find that, when
\begin{align}
\alpha_{k}=\frac{2\pi}{\lambda}(k-1)Md\sin{\theta_0}.
\end{align}
and
\begin{align}
\alpha_{m}=\frac{2\pi}{\lambda}(m-1)d\sin{\theta_0}.
\end{align}
we obtain the maximum power $P_r$. Because of APA, the number of blocks $B$  should be chosen  to be $Q+1$. Thus,  the computational amount of the proposed method in the subsection is
\begin{align}\label{complexHADPA}
C_{HADPA}=O(L(Q+1)(K+M))
\end{align}
FLOPs.
\subsection{Proposed Low-Complexity HDAPA DOA Estimator}\label{S_HDAPA}
In this subsection, we will provide another lower-complexity HPA alternative scheme  with a reverse PA order: DPA, and APA. Firstly, we use the first block of data to perform the DPA by exhaustive linear search. Once we find the feasible set of optimal directions where some pseudo-solutions are included and the number of all solutions are $M$. Secondly, the next $M$ blocks of data are utilized to perform APA.  This means the total number of blocks for PA is  $B=M+1$. Given the initial phases of all analog phase shifters are zeros, the discrete output summation signal of the $k$th subarray corresponding to block $b=1$ is
\begin{align}\label{y_k}
{y}_{k}^{1}(n)&=\mathbf{v}_{A,k}\mathbf{a}_k(\theta_0)s(n)+w_k^1(n)\\\nonumber
&=\frac{1}{\sqrt{M}}[1,~1,\cdots,~1]\mathbf{a}_k(\theta_0)s(n)+w_k^1(n)\\\nonumber
&=\frac{1}{\sqrt{M}}s(n)e^{j\frac{2\pi}{\lambda}(k-1)Md\sin{\theta_0}}\times g(\theta_0)+w_k^1(n),
\end{align}
where
\begin{align}
g(\theta_0)&=\sum_{m=1}^{M}e^{j\frac{2\pi}{\lambda}(m-1)d\sin{\theta_0}}\\\nonumber
&=\frac{1-e^{j\frac{2\pi}{\lambda}Md\sin{\theta_0}}}{1-e^{j\frac{2\pi}{\lambda}d\sin{\theta_0}}},
\end{align}
which are used as the input of digital beamformer in Fig.~1. After passing through DPA, we have
\begin{align}
{r}^1(n)&=\sum_{k=1}^{K}e^{-j\alpha_{k}}{y}_{k}^1(n)\\\nonumber
&=\frac{g(\theta_0)}{\sqrt{M}}s(n)\sum_{k=1}^{K}e^{j(\frac{2\pi}{\lambda}(k-1)Md\sin{\theta_0}-\alpha_{k})}\\\nonumber
&+\sum_{k=1}^{K}e^{-j\alpha_k}w^{1}_{k}(n),
\end{align}
which could be stored in memory at reciever. Furthermore, (\ref{P_r_HADPA}) is represented as
\begin{align}\label{DPA-LS}
{P}^1_r(\hat{\theta}_d)&=\frac{1}{LN^2}\sum_{n=1}^{L}[{r}^1(n){r}^1(n)^*]\\\nonumber
&=\frac{1}{LN^2}\sum_{n=1}^{L}[\big(\frac{g(\theta_0)}{\sqrt{M}}s(n)\sum_{k=1}^{K}e^{j(\frac{2\pi}{\lambda}(k-1)Md\sin{\theta_0}-\alpha_{k})}\\\nonumber
&+\sum_{k=1}^{K}e^{-j\alpha_k}w^{1}_{k}(n)\big)\times\big(\frac{g(\theta_0)}{\sqrt{M}}s(n)\times\\\nonumber
&\sum_{k=1}^{K}e^{j(\frac{2\pi}{\lambda}(k-1)Md\sin{\theta_0}-\alpha_{k})}+\sum_{k=1}^{K}e^{-j\alpha_k}w^{1}_{k}(n)\big)^*],
\end{align}
where
\begin{align}
\alpha_{k}=\frac{2\pi}{\lambda}(k-1)Md\sin{\hat{\theta}_d},
\end{align}
where the angle $\hat{\theta}_d$ is chosen from the angle set $\Theta$. Due to DB, the stepsize $\Delta\theta$ could be set to arbitrarily small. It is assumed that the optimal direction $\hat{\theta}_d$ is attained by an exhaustive linear search over the set $\Theta$. Clearly, $\hat{\theta}_d$ satisfies the following approximate identity
\begin{align}\label{PA_Identity}
\frac{2\pi}{\lambda}(k-1)Md\sin{\theta_0}-\underbrace{\frac{2\pi}{\lambda}(k-1)Md\sin{\hat{\theta}_d}}_{\alpha_{k}}=2i\pi,
\end{align}
where $k\in S_K=\left\{0,~1,~\cdots,K-1\right\}$, and $i\in S_M=\left\{0,~1,~\cdots,~M-1\right\}$. From (\ref{PA_Identity}), we can obtain the set of $M$ feasible solutions for the estimated emitter direction as follows
\begin{align}\label{theta_d_est_set}
\hat{\Theta}_{d}=\left\{\hat{\theta}_{d,0}, \hat{\theta}_{d,1},\cdots,\hat{\theta}_{d,M-1}\right\}.
\end{align}
The $M$ estimation angles in the above equation are substituted into
(\ref{a_m}) which produce $M\times{M}$ matrix $\mathbf{A}_{m}$, i.e.,
\begin{align}
\mathbf{A}_{m}=
\begin{bmatrix}
\alpha_{1,0} & \alpha_{1,1} & \cdots & \alpha_{1,M-1}\\
\alpha_{2,0} & \alpha_{2,1} & \cdots & \alpha_{2,M-1}\\
\vdots & \vdots & \ddots & \vdots\\
\alpha_{M,0} & \alpha_{M,1} & \cdots & \alpha_{M,M-1},
\end{bmatrix}
\end{align}
where $\alpha_{m,i}$ corresponds to $\hat{\theta}_{d,i}$ according to (\ref{a_m}), i.e.,
\begin{align}
\alpha_{m,i}=\frac{2\pi}{\lambda}(m-1)d\sin{\hat{\theta}_{d,i}},
\end{align}
and
\begin{align}
\alpha_{k,i}=\frac{2\pi}{\lambda}(k-1)Md\sin{\hat{\theta}_{d,i}}.
\end{align}
We substitute each column of the above two equations into (\ref{P_r_HADPA}) which will bring $M$ ${P}^b_r(\hat{\theta}_d)$s. At last, we determine $\hat{\theta}_0$ which yields the maximum value of ${P}^b_r(\hat{\theta}_d)$. In the same manner as shown in (\ref{complexHADPA}), the computational amount of the proposed HDAPA is
\begin{align}\label{complexHDAPA}
C_{HDAPA}=O(L(Q+1)K + LM^2)
\end{align}
FLOPs. In particular, we need to clarify what the main differences are between two steps APA and DPA in HADPA and HDAPA. Since APA operates in the analog domain, each-step search corresponding to one bin needs one new block of data because analog signal cannot be stored before ADC operation in Fig.~1. In other words, if the search interval of direction angle is divided into $Q$ bins, then APA requires $Q+1$ blocks of data to complete an exhaustive linear search over the total search range.  Conversely, for the case of DPA, the sampled and quantized signal can be saved in memory. Only one block of data is required to complete an exhaustive linear search over the interval direction angle $[-\pi/2, \pi/2]$.  This means that DPA has a shorter  delay and length of receive data compared with APA. This is the benefit from DPA.

\section{Proposed Low-complexity Hybrid Root-MUSIC-HDAPA Estimator and Hybrid CRLB}\label{S4}
\begin{figure}[tp]
\centering
\includegraphics[width=0.5\textwidth]{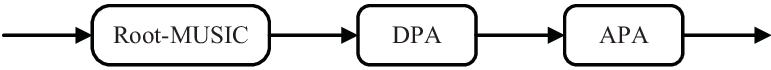}\\
\caption{Schematic diagram for Root-MUSIC-HDAPA.}\label{Root-MUSIC-HDAPA}
\end{figure}

In Section \ref{S3}, we present how to estimate DOA from the aspect of pure linear search. Below, we will use the concept of spatial spectral estimation method to estimate DOA by Root-MUSIC criteria with the aid of HDAPA in Section \ref{S_RM_HDAPA}, which will achieve a faster estimation speed  compared to those methods based on pure linear search. Fig.~\ref{Root-MUSIC-HDAPA} briefly describe the schematic diagram of the proposed Root-MUSIC-HDAPA.
\subsection{Proposed Root-MUSIC-HDAPA DOA Estimator}\label{S_RM_HDAPA}
Here, each subarray will be still viewed as a large virtual antenna, initially, like HDAPA, assume all phases of analog beamforming vector $\mathbf{v}_{A, k}$ are equal to zeros, i.e.,
\begin{align}
\mathbf{v}_{A, k}=\frac{1}{\sqrt{M}}[1,\cdots,1]^{T}.
\end{align}
According to (\ref{y_k}), the output vector of all subarrays at block $1$ is
\begin{align}\label{Y}
\mathbf{y}^1(n)&=[y_1(n),y_2(n),\cdots,y_K(n)]^{T}\\\nonumber
&=\frac{1}{\sqrt{M}}[1,e^{j\frac{2\pi}{\lambda}Md\sin{\theta_0}},\cdots,e^{j\frac{2\pi}{\lambda}(K-1)Md\sin{\theta_0}}]^T\\\nonumber
&\times{g(\theta_0)s(n)}+[w_1^1(n),\cdots,w_K^1(n)]^{T}\\\nonumber
&=\frac{1}{\sqrt{M}}\mathbf{a}_{M}(\theta_0)g(\theta_0)s(n)+[w_1^1(n),\cdots,w_K^1(n)]^{T}.
\end{align}
where
\begin{align}
\mathbf{a}_{M}(\theta_0)=[1,e^{j\frac{2\pi}{\lambda}Md\sin{\theta_0}},\cdots,e^{j\frac{2\pi}{\lambda}(K-1)Md\sin{\theta_0}}]^T,
\end{align}
$\mathbf{a}_{M}(\theta_0)$ can be viewed as the array manifold vector of the virtual array with each subarray as its virtual antenna elements, and $g(\theta_0)$ is the common factor due to the summation of all elements per subarray. Let us define
\begin{align}
\mathbf{a}_{D}(\theta_0)=g(\theta_0)\mathbf{a}_{M}(\theta_0),
\end{align}
Thus, $\mathbf{y}^1(n)$ in (\ref{Y}) is written as
\begin{align}
\mathbf{y}^1(n)=\frac{1}{\sqrt{M}}\mathbf{a}_{D}(\theta_0)s(n)+[w_1^1(n),\cdots,w_K^1(n)]^{T},
\end{align}

 Now, we adopt the Root-MUSIC algorithm in digital part to estimate DOA. The covariance matrix of the output vector $\mathbf{y}^1(n)$ of virtual antenna array in Fig.~1  is
\begin{align}\label{R-yy-Covar-Mat}
\mathbf{R}_{yy}&=\mathbb{E}[\mathbf{y}\mathbf{y}^{H}]\\\nonumber
&=\mathbf{a}_{D}\mathbf{R}_{ss}\mathbf{a}_{D}^{H}+\mathbf{R}_{ww}\\\nonumber
&=\frac{1}{M}\sigma_s^{2}\|g(\theta_0)\|^2\mathbf{a}_{M}(\theta_0)\mathbf{a}_{M}^{H}(\theta_0)+\sigma_{w}^{2}\mathbf{I},
\end{align}
where $\sigma_s^2$ represents the variance of the receive signal, which equals the average receive signal power. Furthermore, similar to the conventional Root-MUSIC method, the singular-value decomposition (SVD) of $\mathbf{R}_{yy}$ is expressed as
\begin{align}
\mathbf{R}_{yy}=\left[\mathbf{E}_{S}~\mathbf{E}_{N}\right]\mathbf{\Sigma}\left[\mathbf{E}_{S}~\mathbf{E}_{N}\right]^H,
\end{align}
where $\mathbf{E}_{S}$ denotes the $K\times1$ column vector consisting of the singular vector corresponding to the
largest singular value, the matrix $\mathbf{E}_{N}$ contains the singular vectors corresponding to $K-1$ smallest singular values, and the $K\times K$ diagonal matrix $\mathbf{\Sigma}$ has the following form
\begin{align}
\mathbf{\Sigma}=
\begin{bmatrix}
\sigma_s^2+\sigma_{w}^{2} & 0 & \cdots & 0\\
0 & \sigma_{w}^{2} & \cdots & 0\\
\vdots & \vdots & \ddots & \vdots\\
0 & 0 & \cdots & \sigma_{w}^{2}
\end{bmatrix}.
\end{align}
Using the definition of  pseudo spectrum of MUSIC algorithm in \cite{Tuncer}, we have the corresponding  pseudo spectrum
\begin{align}\label{MUSIC-PS}
P_{MU}({\theta})&=\frac{1}{\|\mathbf{a}_{D}^{H}({\theta})\mathbf{E}_{N}\mathbf{E}_{N}^{H}\mathbf{a}_{D}({\theta})\|}\\\nonumber
&=\frac{1}{\|g({\theta})\|^{2}\|\mathbf{a}_{M}^{H}({\theta})\mathbf{E}_{N}\mathbf{E}_{N}^{H}\mathbf{a}_{M}({\theta})\|}.
\end{align}

By maximizing the above $P_{MU}({\theta})$, we have obtain the emitter direction. In general, there are two kinds of ways to estimate the emitter direction: linear search and Root-MUSIC. The latter is attractive due to its low-complexity and near-analytic solution. In what follows, we will design a modified Root-MUSIC algorithm to find the optimal direction in the case of our hybrid structure, which is different from fully-digital structure. Considering that the denominator in the right side of equation (\ref{MUSIC-PS}) is close to zero for $\theta\approx\theta_0$, we define the polynomial equation
\begin{align}\label{D_theta_phi_z}
f_{\theta}({\theta})&=g^{H}({\theta})\mathbf{a}^{H}_{M}({\theta})\mathbf{E}_{N}\mathbf{E}_{N}^{H}\mathbf{a}_{M}({\theta})g({\theta})\\\nonumber
&=\frac{2-e^{-j\frac{2\pi}{\lambda}Md\sin{{\theta}}}-e^{j\frac{2\pi}{\lambda}Md\sin{{\theta}}}}{2-e^{-j\frac{2\pi}{\lambda}d\sin{{\theta}}}-e^{j\frac{2\pi}{\lambda}d\sin{\hat{\theta}}}}\times\\\nonumber
&\sum_{m=1}^{K}\sum_{n=1}^{K}e^{-j\frac{2\pi}{\lambda}Md(m-1)\sin{{\theta}}}\mathbf{C}_{mn}e^{j\frac{2\pi}{\lambda}Md(n-1)\sin{{\theta}}}\\\nonumber
&\triangleq{f_z(z)}\triangleq{f_{\phi}(\phi)}=0,
\end{align}
where $\mathbf{C}=\mathbf{E}_{N}\mathbf{E}_{N}^{H}$, $\mathbf{C}_{mn}$ is the element in the $n$th column of the $m$th row of $\mathbf{C}$,
\begin{align}\label{z-expr-theta}
z=e^{j\frac{2\pi}{\lambda}Md\sin{{\theta}}},
\end{align}
and
\begin{align}\label{phi-expr-theta}
 \phi=\frac{2\pi}{\lambda}Md\sin{\theta},
\end{align}
then (\ref{D_theta_phi_z}) is rewritten in the simple form
\begin{align}\label{D_z}
f_z(z)&=\frac{2-z^{-1}-z}{2-z^{-\frac{1}{M}}-z^{\frac{1}{M}}}\sum_{m=1}^{K}\sum_{n=1}^{K}z^{-(m-1)}\mathbf{C}_{mn}z^{(n-1)}\\\nonumber
&=0.
\end{align}

Observing the above polynomial equation, we find its highest degree is  $2K-2$. This means that this equation has $2K-2$ roots. When $z_0$ is a root of $f_{z}(z)$, $1/{z_0^{*}}$ is its root as well. Now, we define the set of its $2K-2$ roots as follows
\begin{align}
Z_{RM}=\left\{z_i,~i\in\left\{1,2,\cdots,2K-2\right\}\right\},
\end{align}
which yields the set of associated emitter directions as follows

\begin{align}
\hat{\Theta}_{RM}=\left\{\hat{\theta}_{i},~i\in\left\{1,2,\cdots,2K-2\right\}\right\}
\end{align}
where
\begin{align}
\hat{\theta}_{i}=\arcsin\left(\frac{\lambda\arg{z_i}}{2\pi{Md}}\right).
\end{align}
Now, we use the digital beamformer (\ref{DPA-LS}) to keep the true optimal solution by deleting other $2K-3$  pseudo-solutions in $\hat{\Theta}_{RM}$ , which is formulated as the following optimization problem
\begin{equation}\label{max-Rx-power-theta}
 \hat{\theta}_{RM-DPA}=\mathop{\argmax}_{\hat{\theta}_d\in \hat{\Theta}_{RM}}{\tilde{P}^1_r(\hat{\theta}_d)},
\end{equation}
which yields
\begin{align}
  \hat{\phi}_{RM-DPA}=\frac{2\pi}{\lambda}Md\sin{\hat{\theta}_{RM-DPA}},
\end{align}
and
\begin{align}\label{z-expr}
 \hat{z}_{RM-DPA}=e^{j\frac{2\pi}{\lambda}Md\sin{{\hat{\theta}_{RM-DPA}}}}
\end{align}
from (\ref{D_theta_phi_z}). Observing (\ref{D_theta_phi_z}), (\ref{z-expr-theta}),~and (\ref{phi-expr-theta}),~it is evident that the function $f_{\phi}(\phi)$ is a periodic function of $\phi$ with period $2\pi$.
In other words,~$f_{\phi}(\hat{\phi}_{RM-DPA,i})=f_{\phi}(\hat{\phi}_{RM-DPA}+2i\pi)$,~and
$z_{RM-DPA,~i}=e^{j\phi_{RM-DPA,i}}$,~for $i\in\{0,1,\cdots,M-1\}$,~form all feasible solutions to (\ref{D_z}). Thus, we have the extended feasible set as follows
\begin{align}¡¢
\hat{\Theta}_{RM-DPA}=\left\{\hat{\theta}_{RM-DPA,~i},~i\in\left\{0,1,\cdots,M-1\right\}\right\},
\end{align}
where
\begin{align}\label{theta}
\hat{\theta}_{RM-DPA,i}=\arcsin\left(\frac{\lambda(\arg{\hat{z}_{RM-DPA}}+2\pi{i})}{2\pi{Md}}\right).
\end{align}

Considering the objective function in (\ref{max-Rx-power-theta}) is also a periodic function of $\phi$ with period $2\pi$. Excluding the pseudo-solution in feasible set $\hat{\Theta}_{RM-DPA}$ of solutions  requires APA. Before ADC, it is impossible to store the analog signal, then we need to use the next new $M$ blocks of signals.

Therefore, similar to the HADPA, we compute the set of all ${P}^b_r(\hat{\theta}_d)$s in (\ref{DPA-LS}) corresponding to all $M$ phases in $\hat{\Theta}_{RM-DPA}$ as follows
\begin{equation}
S_P=\left\{P_r^1\left(\hat{\Theta}_{RM-DPA,0}\right),~\cdots, P_r^M\left(\hat{\Theta}_{RM-DPA,~M-1}\right)\right\}.
\end{equation}
The value of emitter direction $\hat{\theta}_0$ associated with the largest element in set $S_P$ is the resulting estimated direction angle. This completes the estimate process of the proposed Root-MUSIC-HDAPA scheme.
\subsection{Hybrid CRLB}
To evaluate the proposed HDAPA and Root-MUSIC-HDAPA methods above, the CRLB for hybrid structure, based on (\ref{R-yy-Covar-Mat}), is derived in Appendix A and is described in the following theorem.

\textbf{Theorem 1:}  For the HAD beamforming structure in Fig.~1, with single emission source and ULA, the variance of unbiased DOA estimator is lower bounded by the following hybrid CRLB
 \begin{align}
\sigma^2_{\theta}\ge\frac{1}{N_s}\mathrm{Tr}\left(\mathbf{F}^{-1}\right)
\end{align}
where
\begin{align}
\mathbf{F}&=\frac{8\pi^2\cos^2{\theta}SNR^2}{\lambda^2M(M+KSNR\|g(\theta)\|^2)}\\\nonumber
&\times\Big(\frac{\|g(\theta)\|^4}{6}M^2K^2(K-1)(2K-1)d^2\\\nonumber
&-\frac{\|g(\theta)\|^4}{4}M^2K^2(K-1)^2d^2\\\nonumber
&+\frac{\|g(\theta)\|^2MK}{M+KSNR\|g(\theta)\|^2}\|\eta\|^2\\\nonumber
&+\frac{MK^2}{M+KSNR\|g(\theta)\|^2}{Re}\{g^2(\theta)\eta\}\Big).
\end{align}
\emph{Proof:} See Appendix A.\hfill$\blacksquare$
\subsection{Complexity Analysis and Comparison}
According to (\ref{complexHDAPA}), the computational amount of the Root-MUSIC-HDAPA is
\begin{align}
 C_{RM-HDAPA}&=O(K^2L+(2(K-1))^3\\\nonumber
 &+L((2K-2)K+M^2))
\end{align}
FLOPs. Regardless of computational complexity, APA and HADPA require more time-domain blocks to implement one-time phase alignment so as to estimate the high-resolution DOA compared with HDAPA and Root-MUSIC-HDAPA. The required numbers of time blocks for Root-MUSIC-HDAPA, HDAPA, HADPA, and APA are as follows: $M+1$, $M+1$, $Q+1$, and $Q+1$, respectively.  Obviously, the numbers of time blocks for Root-MUSIC-HDAPA and HDAPA are $M+1$, independent of stepsize, smaller than $Q+1$, i.e., the number of HADPA and APA depending on stepsize. In general, $M$ is far smaller than $Q$.  Actually, the number of time blocks has a profound impact on the computational complexity as listed in Table I. Reversely, as shown in Table I, the computational complexity of each method  is a linear function of the corresponding number of time blocks.
\newcommand{\tabincell}[2]{\begin{tabular}{@{}#1@{}}#2\end{tabular}}
\begin{table}[h]
\centering
\begin{tabular}{cc}
\hline
Algorithms & Complexity~~~~\\
\hline
Conventional APA  & $O((Q+1)LN)$~~~~\\
\hline
Proposed HADPA & $O((Q+1)L(K+M))$~~~~\\
\hline
Proposed HDAPA &  $O((Q+1)LK+LM^2)$~~~~\\
\hline
Proposed Root-MUSIC-HDAPA & \tabincell{c}{$O(K^2L+(2(K-1))^3$ \\ $+L((2K-2)K+M^2))$}\\
\hline
\end{tabular}
\caption{Complexity Comparison}
\end{table}

\section{Simulation Results}\label{S5}
In this section, we present simulation results to demonstrate the performance of the three DOA estimators proposed by us: HADPA, HDAPA, and Root-MUSIC-HDAPA. Simulation parameters are chosen as follows: the direction of emitter $\theta_0=41.177^\circ$, $L=32$, and $M\in\left\{1,~2,~4,~8\right\}$. In medium-scale and large-scale MIMO scenarios, the number $N$ of antennas at receive array is set to $32$ and $128$, respectively.

\begin{figure}[tp]
\centering
\includegraphics[width=0.5\textwidth]{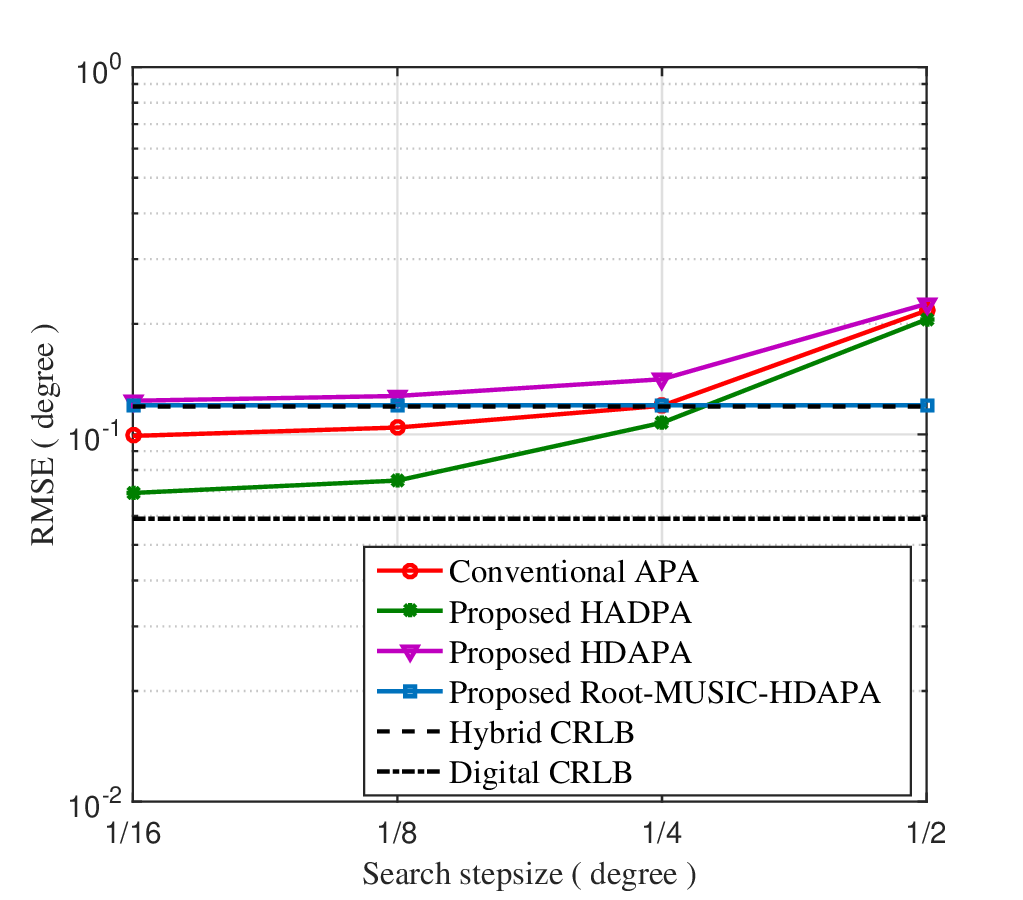}\\
\caption{RMSE versus stepsize of the proposed $3$ methods for $K=16$, $N=32$, and $SNR=0dB$.}\label{search32}
\end{figure}

\begin{figure}[tp]
\centering
\includegraphics[width=0.5\textwidth]{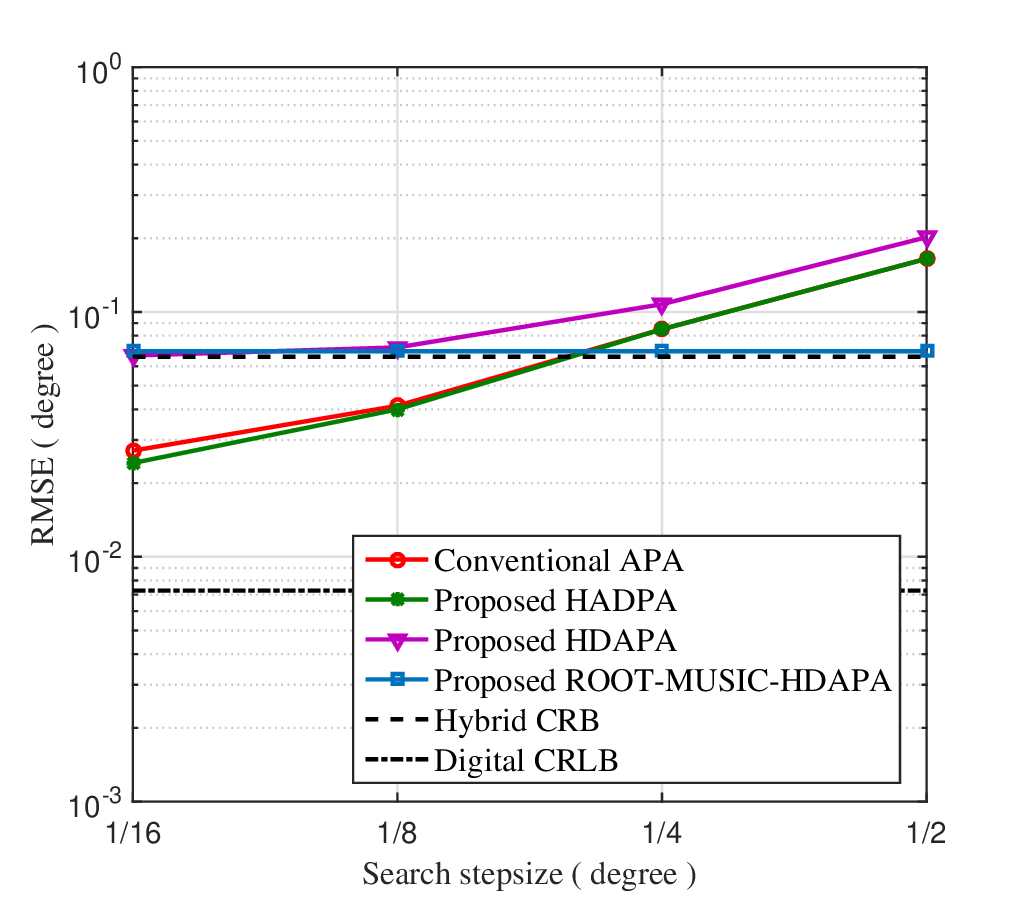}\\
\caption{RMSE versus stepsize of the proposed $3$ methods for $K=16$, $N=128$, and $SNR=0dB$.}\label{search128}
\end{figure}

Firstly, Fig.~\ref{search32} and Fig.~\ref{search128} plot the curves of root mean squared error (RMSE) versus stepsize of the four DOA estimators  APA, HADPA, and HDAPA in Section III, and the proposed Root-MUSIC-HDAPA in Section IV for different values of $N$: 32 (medium-scale) and 128 (large-scale), respectively. Here, $K$, the number of subaarays, is set to $16$, and SNR is equal to $0dB$. It is seen from the two figures that the RMSE performance of all three methods of linear searching proposed in Section III improve as stepsize decreases. In particular, when stepsize is small enough, APA and HADPA will be closer to the fully-digital CRLB while the proposed HDAPA and Root-MUSIC-HDAPA can converge to the hybrid CRLB. In large scale case, we observe that, when stepsize exceeds $0.25^\circ$, the proposed Root-MUSIC-HDAPA performs better than two pure linear searching algorithms: APA, and HADPA. Conversely, it is worsen than  APA, and HADPA. However, a small stepsize means high complexity. In what follows, we will compare their complexity. The proposed Root-MUSIC-HDAPA  owns an extremely lower computational complexity than other methods. Thus, below, we will make a deep and extensive investigation on the proposed Root-MUSIC-HDAPA.
\begin{figure}[h]
\centering
\includegraphics[width=0.5\textwidth]{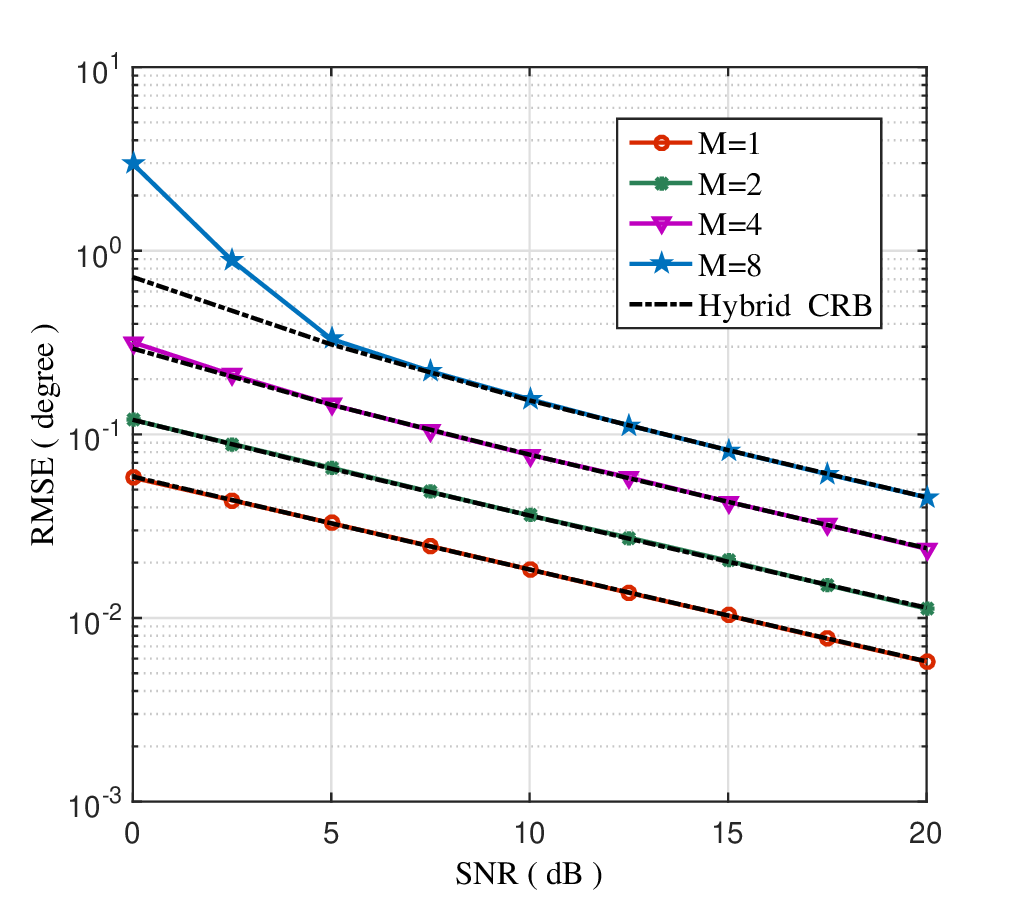}\\
\caption{RMSE versus SNR of the proposed Root-MUSIC-HDAPA for $M\in\left\{1,~2,~4,~8\right\}$, and $N=32$.}\label{N32_M}
\end{figure}

Fig.~\ref{N32_M} shows the performance curves of RMSE versus SNR of  the proposed Root-MUSIC-HDAPA algorithm with $N=32$, $L=32$, and $M\in\left\{1,~2,~4,~8\right\}$, where the corresponding CRLBs are used as a performance benchmark. From Fig.~\ref{N32_M}, it is obvious that the proposed  Root-MUSIC-HDAPA method can achieve the corresponding CRLBs as SNR exceeds a fixed threshold. For example, $M=8$ and $N=32$,  the proposed method can reach the CRLB curve when SNR is larger than 5dB.
Also, we find that as $M$ increases, the RMSE performance of the proposed method degrades gradually, and the corresponding CRLB value increases.
\begin{figure}[h]
\centering
\includegraphics[width=0.5\textwidth]{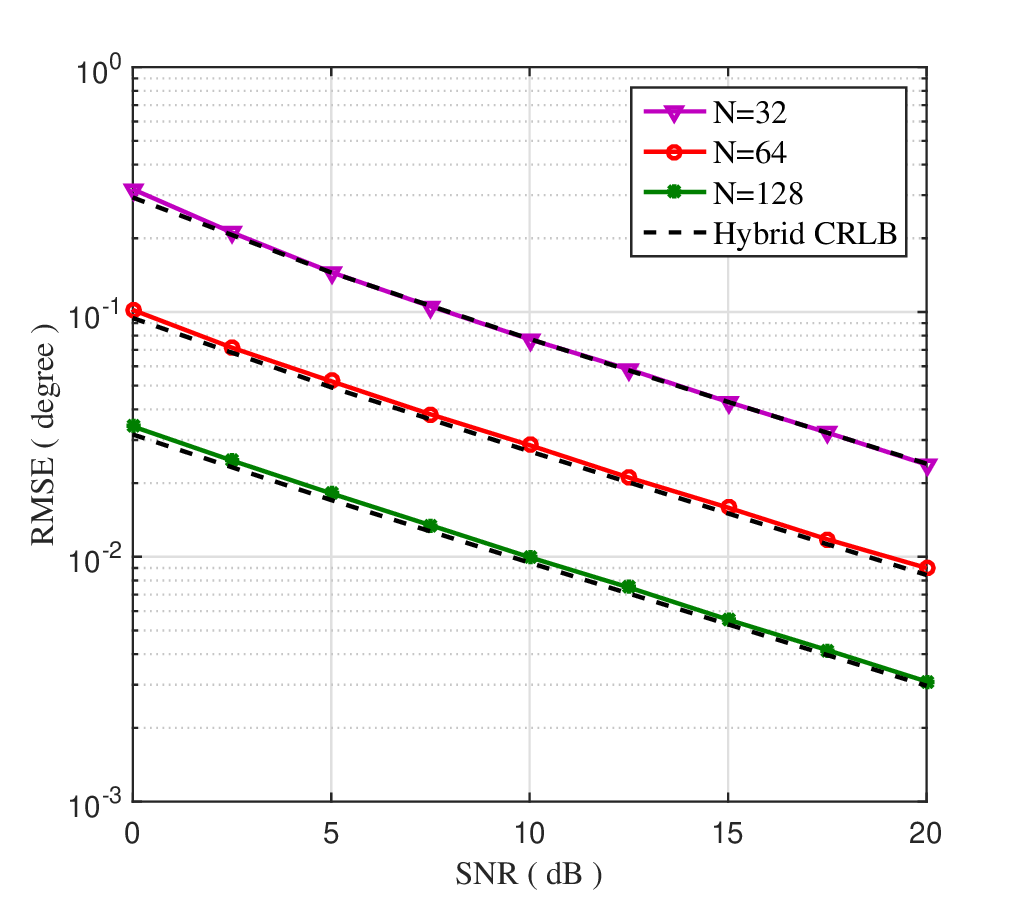}\\
\caption{RMSE of the estimation errors for $M=4$ with $N=32,64,128$.}\label{fixedM4}
\end{figure}

To observe the impact of the total number $N$ of array antennas on the proposed Root-MUSIC-HDAPA scheme,  in Fig.~\ref{fixedM4}, we change the value of $N$ from 32 to 128, given fixed $L=32$, and $M=4$. Similar to Fig.~\ref{N32_M}, Fig.~\ref{fixedM4} still plots the RMSE verus SNR curves of the proposed Root-MUSIC-HDAPA method. From this figure, we obtain the same performance trend as Fig.~\ref{N32_M}. Particularly, we note that, as the total number of antennas $N$ increases, the accuracy of the proposed algorithm  improves accordingly.

\begin{figure}[tp]
\centering
\includegraphics[width=0.5\textwidth]{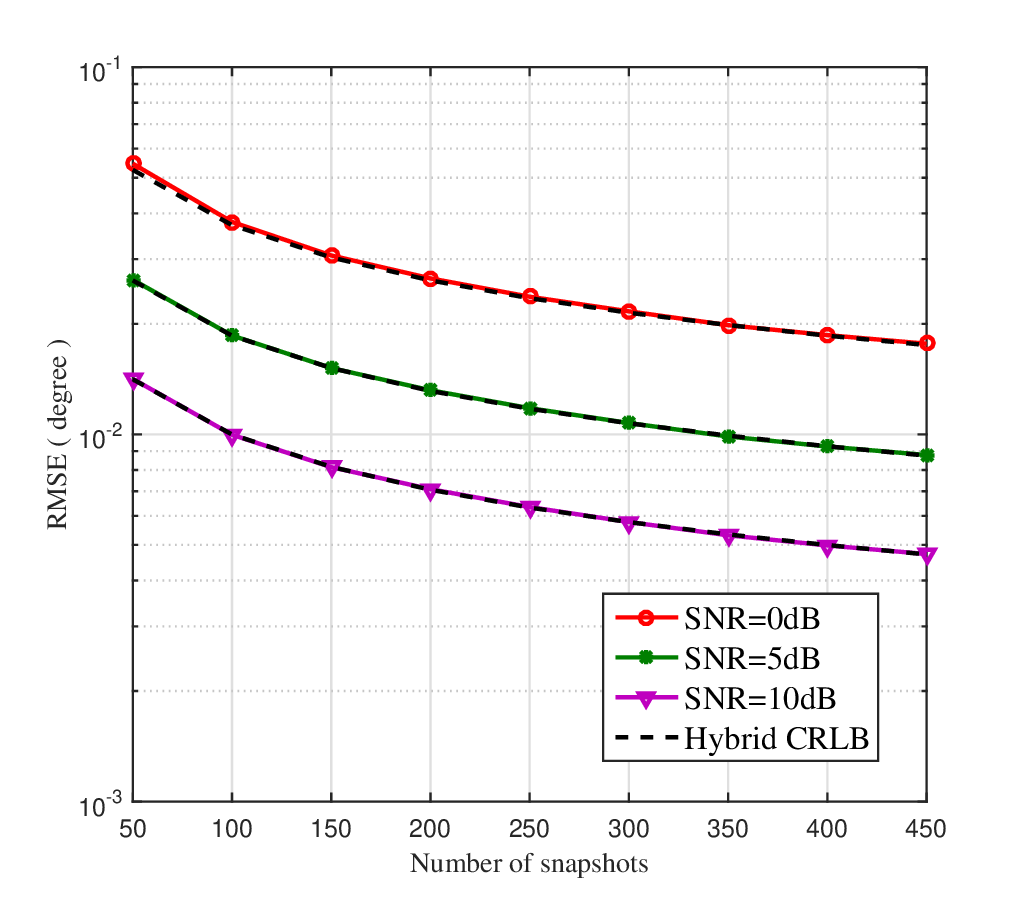}\\
\caption{RMSE comparison for proposed Root-MUSIC-HDAPA  with $N=128$ and $M=8$}\label{snapshot}
\end{figure}

Fig.\ref{snapshot} illustrates the RMSE performance  versus the number $L$ of snapshots for three different values of SNR: 0dB, 5dB, and 10dB. Regardless of the value of SNR and the number of snapshots/sampling points, the RMSE performance  will always reach the corresponding CRLB. Additionally, as the number $L$ of snapshots increases, the RMSE  performance becomes better and better.
\begin{figure}[h]
\centering
\includegraphics[width=0.5\textwidth]{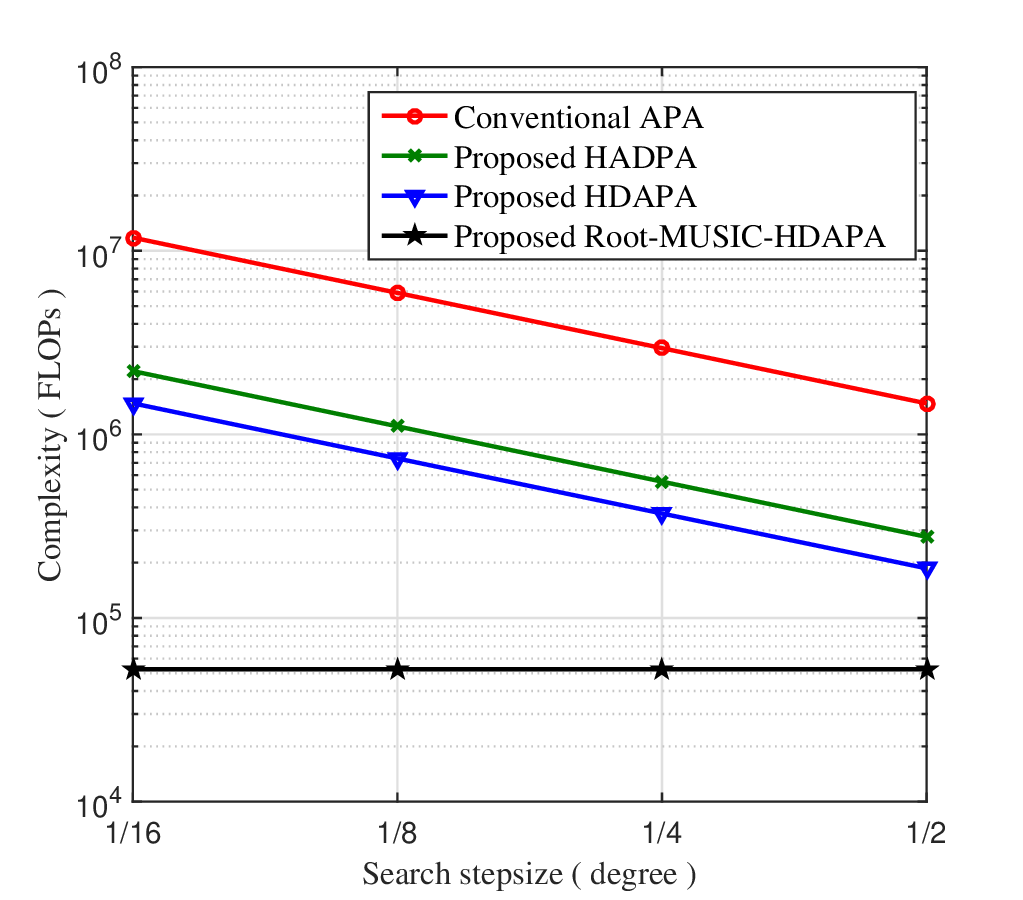}\\
\caption{Complexity comparison for proposed $4$ methods with $N=128$ and $M=8$}\label{complexQ}
\end{figure}

\begin{figure}[tp]
\centering
\includegraphics[width=0.5\textwidth]{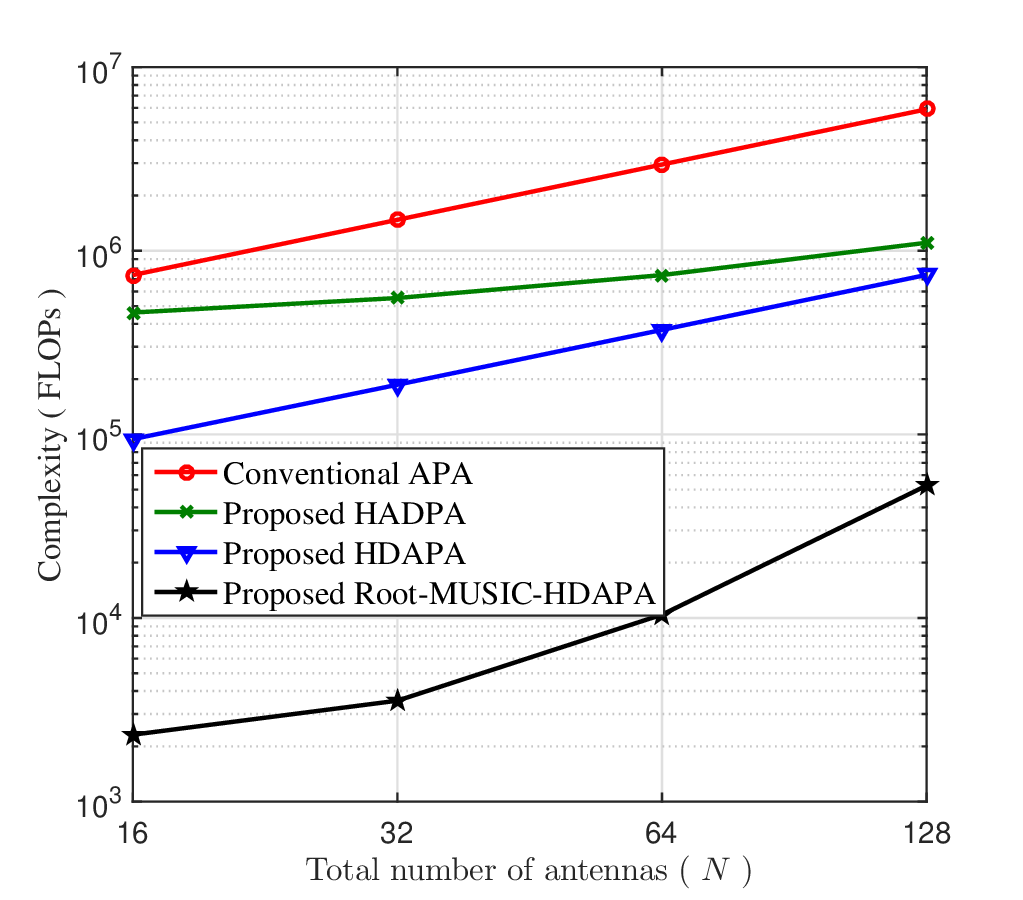}\\
\caption{Complexity comparison for proposed $4$ methods with $\Delta\theta=\frac{1}{8}^\circ$ and $M=8$}\label{complexN}
\end{figure}

The computational complexity of all methods including conventional APA, and the three proposed methods are listed in Table I. As $N=KM$ and $Q$ tends to large-scale, the first three methods APA, HADPA, and HDPAP algorithm has much higher complexity than the last one Root-MUSIC-HDAPA, where a large value of $Q$  leads to a high-resolution DOA estimation precision for the first three methods.  In the last method, the set of linear searching directions is fixed and independent of resolution requirement.  This is why it has the lowest-complexity one among the four methods. To further assess their complexity relationship, their complexity curves are also plotted in Fig.\ref{complexQ} and Fig.\ref{complexN}.

Fig.~\ref{complexQ} illustrates the curves of complexity versus search stepsize with $N=128$, $M=8$ and $L=32$. We note that the four methods have the following decreasing order in complexity: APA, HADPA, HDAPA, and Root-MUSIC-HDAPA. Clearly, the proposed Root-MUSIC-HDAPA achieve the lowest complexity among all four methods. More importantly, even compared to HDAPA, it is still lower by near-an-order-of-magnitude.

Fig.~\ref{complexN} shows the curves of complexity versus the number of antennas with $M=8$,  $L=32$, $\Delta\theta=0.125^\circ$, and $N\in\{16,32,64,128\}$. From this figure, it is seen  that as the number of total antennas increases, the complexity of all algorithms increases rapidly.  However, the proposed Root-MUSIC-HDAPA is still the lowest-complexity one among them and  lower than them by near-an-order-of-magnitude compared with the second low-complexity HDAPA.

In summary, the proposed Root-MUSIC-HDAPA and HDAPA can achieve the hybrid-structure CRLB with dramatically lower complexity than HADPA and APA. Due to linear search in HDAPA, the latter complexity is subtatantially higher than the former. The proposed HADPA method can reach the fully-digital CRLB with the third low-complexity. However, the conventional APA needs an extremely high complexity to attain the fully-digital CRLB.

\section{Conclusion}\label{S6}
In this paper, based on the hybrid structure,  we proposed three DOA estimators: HADPA, HDAPA, and Root-MUSIC-HDAPA. The first two schemes are the type of pure linear searching. The last one is a hybrid method consisting of two steps: approximately closed-form solution in the first step and linear searching over a set of limited finite directions predetermined by the first step in the second step. This leads to an extremely computational complexity for the last one, which is significantly lower  than those of APA, HADPA, and HDAPA. By simulation and analysis, we find the proposed  Root-MUSIC-HDAPA and HDAPA can reach the hybrid CRLB with very low-complexity, and the HADPA and APA can achieve the full-digital CRLB with far higher complexity than the former two methods. In summary, the proposed Root-MUSIC-HDAPA and HDAPA methods strike an excellent balance among accuracy, complexity and number of time blocks. This makes them attractive for the future applications of measuring DOA in IoT, UAV, satellite communications, WSNs, and 5G and beyond.

\appendices
\section{Derivation of CRLB for Hybrid Structure}
In the following, the CRLB for hybrid analog-and-digital receive array is derived. In accordance with \cite{Moon},  the Fisher information matrix (FIM) $\mathbf{F}$ for hybrid structure is given by
\begin{align}\label{Fisher-Mat}
\mathbf{F}=\mathrm{Tr}\left\{\mathbf{R}_{yy}^{-1}\frac{\mathrm{d}\mathbf{R}_{yy}}{\mathrm{d}\theta}\mathbf{R}_{yy}^{-1}\frac{\mathrm{d}\mathbf{R}_{yy}}{\mathrm{d}\theta}\right\},
\end{align}
where the covariance matrix $\mathbf{R}_{yy}$ is defined  in (\ref{R-yy-Covar-Mat}), which is rewritten as
\begin{align}
\mathbf{R}_{yy}=\gamma\mathbf{V}_A^H\mathbf{a}(\theta)\mathbf{a}^{H}(\theta)\mathbf{V}_A+\mathbf{I},
\end{align}
where $\sigma^2_w=1$ and $\gamma=\sigma^2_s/\sigma^2_w$.
To calculate the Fisher information matrix (\ref{Fisher-Mat}), we first attain the term
\begin{align}
\frac{\mathrm{d}\mathbf{R}_{yy}}{\mathrm{d}\theta}&=\gamma\mathbf{V}_A^{H}(\mathbf{\dot{a}}(\theta)\mathbf{a}^{H}(\theta)+\mathbf{a}(\theta)\mathbf{\dot{a}}^{H}(\theta))\mathbf{V}_A,
\end{align}
where  $\mathbf{a}(\theta)$ is the subarray manifold given in (\ref{d}), and its  derivative with respect to $\theta$ is
\begin{align}\label{a-theta}
\mathbf{\dot{a}}(\theta)=j\frac{2\pi}{\lambda}\cos{\theta}\mathbf{D}\mathbf{a}(\theta),
\end{align}
where
\begin{align}
\mathbf{D}=\text{diag}\{d_1,d_2,\cdots,d_N\}.
\end{align}
In the following, for convenience of deriving,  $\mathbf{a}(\theta)$ and $\mathbf{\dot{a}}(\theta)$ are abbreviated as $\mathbf{a}$ and $\mathbf{\dot{a}}$, respectively.
Therefore,
\begin{align}\label{F-expre}
\mathbf{F}&=\gamma^{2}\mathrm{Tr}\{\mathbf{R}_{yy}^{-1}\mathbf{V}_{A}^{H}(\mathbf{\dot{a}}\mathbf{a}^{H}+\mathbf{a}\mathbf{\dot{a}}^{H})\mathbf{V}_{A}\\\nonumber
&\times\mathbf{R}_{yy}^{-1}\mathbf{V}_{A}^{H}(\mathbf{\dot{a}}\mathbf{a}^{H}+\mathbf{a}\mathbf{\dot{a}}^{H})\mathbf{V}_{A}\},
\end{align}
which is expanded and combined to form
\begin{align}\label{F-expre-0}
\mathbf{F}&=\gamma^{2}\Big[\underbrace{\left(\mathbf{a}^{H}\mathbf{V}_{A}\mathbf{R}_{yy}^{-1}\mathbf{V}_{A}^{H}\mathbf{\dot{a}}\right)^2}_{F_1}\\\nonumber
&+2\underbrace{\left(\mathbf{a}^{H}\mathbf{V}_{A}\mathbf{R}_{yy}^{-1}\mathbf{V}_{A}^{H}\mathbf{a}\right)\left(\mathbf{\dot{a}}^{H}\mathbf{V}_{A}\mathbf{R}_{yy}^{-1}\mathbf{W}_{A}^{H}\mathbf{\dot{a}}\right)}_{F_2}\\\nonumber
&+\underbrace{\left(\mathbf{\dot{a}}^{H}\mathbf{V}_{A}\mathbf{R}_{yy}^{-1}\mathbf{V}_{A}^{H}\mathbf{a}\right)^{2}}_{F_3}\Big]=\gamma^2\left(F_1+F_2+F_3\right)
\end{align}
where
\begin{align}
\mathbf{R}_{yy}^{-1}=\mathbf{I}-\frac{1}{\gamma^{-1}+\mathbf{a}^{H}\mathbf{V}_{A}\mathbf{V}_{A}^{H}\mathbf{a}}\mathbf{V}_{A}^{H}\mathbf{a}\mathbf{a}^{H}\mathbf{V}_{A},
\end{align}
and
\begin{align}
&\mathbf{a}^{H}\mathbf{V}_{A}\mathbf{R}_{yy}^{-1}\mathbf{V}_{A}^{H}\mathbf{\dot{\mathbf{a}}}
=\mathbf{a}^H\mathbf{V}_A\mathbf{V}_A^H\dot{\mathbf{a}}\\\nonumber
&-\frac{\mathbf{a}^H\mathbf{V}_A\mathbf{V}_A^H\mathbf{a}\mathbf{a}^H\mathbf{V}_A\mathbf{V}_A^H\dot{\mathbf{a}}}{\gamma^{-1}+\mathbf{a}^H\mathbf{V}_A\mathbf{V}_A^H\mathbf{a}}.
\end{align}
We note that
\begin{align}
\mathbf{V}_{A}\mathbf{V}_A^H
&=\frac{1}{M}\begin{bmatrix}
1 & \cdots & 1 & ~ & ~ & ~ & ~\\
\vdots & ~ & \vdots & ~ & ~ & \mathbf{0} &~\\
1 & \cdots & 1 & ~ & ~ & ~ & ~\\
~ & ~ & ~ & \ddots & ~ & ~ & ~\\
~ & ~ & ~ & ~ & 1 & \cdots & 1\\
~ & \mathbf{0} & ~ & ~ & \vdots & ~ & \vdots\\
~ & ~ & ~ & ~ & 1 & \cdots & 1\\
\end{bmatrix}
&=\frac{1}{M}\mathbf{B},
\end{align}
where $\mathbf{B}$ is the block diagonal matrix which consists of $K$ $M\times{M}$  matrices of all-ones. Let us define $\mathbf{a}^{H}\mathbf{B}\mathbf{a}=\Gamma$.
In the first step, we compute the expression of $F_1$. Then
\begin{align}\label{F-1-expres}
F_1&=\left(\mathbf{a}^{H}\mathbf{V}_{A}\mathbf{R}_{yy}^{-1}\mathbf{V}_{A}^{H}\mathbf{\dot{a}}\right)^2\\\nonumber
&=-\frac{4\pi^2\cos^2{\theta}}{M^2\lambda^2}\left(1-\frac{\Gamma}{\frac{M}{\gamma}+\Gamma}\right)^2\left(\mathbf{a}^{H}\mathbf{B}\mathbf{D}\mathbf{a}\right)^2.
\end{align}
Obviously, to obtain the detailed expression of $F_1$, we have to know $\mathbf{a}^{H}\mathbf{B}\mathbf{D}\mathbf{a}$ in advance. Making  a utilization of the Kronecker product in \cite{Moon}, matrix $\mathbf{B}$ could be represented as
\begin{align}
\mathbf{B}=\mathbf{I}_{K}\otimes{\mathbf{E}_{M}},
\end{align}
where $\mathbf{E}_{M}$ stands of the  $M\times{M}$ matrix of all ones. Then, the array manifold $\mathbf{a}$ is simplified as
\begin{align}
\mathbf{a}=\mathbf{a}_{D}\otimes{\mathbf{a}_{A}},
\end{align}
where
\begin{align}
\mathbf{a}_{D}=[1,e^{j\frac{2\pi}{\lambda}Md\sin{\theta}},\cdots,e^{j\frac{2\pi}{\lambda}(K-1)Md\sin{\theta}}]^T,
\end{align}
and
\begin{align}
\mathbf{a}_{A}=[1,e^{j\frac{2\pi}{\lambda}d\sin{\theta}},\cdots,e^{j\frac{2\pi}{\lambda}(M-1)d\sin{\theta}}]^T.
\end{align}
According to the definition of $\mathbf{D}$ in (39), we have
\begin{align}
\mathbf{D}=\mathbf{I}_{K}\otimes{\mathbf{D}_A}+{\mathbf{D}_D}\otimes{\mathbf{I}_{M}},
\end{align}
where
\begin{align}
\mathbf{D}_A=\begin{bmatrix}
d_1 & 0 & \cdots & 0\\
0 & d_2 & \cdots & 0\\
\vdots & \vdots & \ddots & \vdots\\
0 & 0 & \cdots & d_M\\
\end{bmatrix},
\end{align}
and
\begin{align}
\mathbf{D}_D=
\begin{bmatrix}
0 & 0 & \cdots & 0\\
0 & d_{M+1} & \cdots & 0\\
\vdots & \vdots & \ddots & \vdots\\
0 & 0 & \cdots & d_{(K-1)M+1}\\
\end{bmatrix}.
\end{align}
Therefore,
\begin{align}\label{aHBDa-0}
\mathbf{a}^H\mathbf{BDa}&=(\mathbf{a}_{D}\otimes{\mathbf{a}_{A}})^H(\mathbf{I}_{K}\otimes{\mathbf{E}_{M}})\\\nonumber
&\times(\mathbf{I}_{K}\otimes{\mathbf{D}_A}+{\mathbf{D}_D}\otimes{\mathbf{I}_{M}})(\mathbf{a}_{D}\otimes{\mathbf{a}_{A}}).
\end{align}
According to basic property of the Kronecker product in \cite{Moon}, i.e.,
\begin{align}
\mathrm{Tr}(A\otimes{B})=\mathrm{Tr}(A)\mathrm{Tr}(B),
\end{align}
$\mathbf{a}^H\mathbf{BDa}$ in (\ref{aHBDa-0}) is written as
\begin{align}\label{aHBDa}
\mathbf{a}^H\mathbf{BDa}&=\mathrm{Tr}(\mathbf{a}_D^H\mathbf{a}_D)\mathrm{Tr}(\mathbf{a}_A^H\mathbf{E}_M\mathbf{D}_A\mathbf{a}_A)\\\nonumber
&+\mathrm{Tr}(\mathbf{a}_D^H\mathbf{D}_D\mathbf{a}_D)\mathrm{Tr}(\mathbf{a}_A^H\mathbf{E}_M\mathbf{a}_A)\\\nonumber
&=Kg(-\theta)\zeta+\sum_{k=1}^{K}\mathbf{D}_{D,k}g(-\theta)g(\theta),
\end{align}
where
\begin{align}
\zeta=\sum_{m=1}^{M}d_{m}e^{j\frac{2\pi}{\lambda}d_m\sin{\theta}}.
\end{align}

Considering $g(-\theta)=g^{H}(\theta)$, inserting the above expression into the right-hand side of $F_1$ in (\ref{F-1-expres}) gives
\begin{align}\label{F-1-expres-1}
F_1&=\left(\mathbf{a}^{H}\mathbf{V}_{A}\mathbf{R}_{yy}^{-1}\mathbf{V}_{A}^{H}\mathbf{\dot{a}}\right)^2\\\nonumber
&=-\frac{4\pi^2\cos^2{\theta}}{M^2\lambda^2}\left(1-\frac{\Gamma}{\frac{M}{\gamma}+\Gamma}\right)^2\\\nonumber
&\times\left(Kg(-\theta)\zeta+\sum_{k=1}^{K}\mathbf{D}_{D,k}\|g(\theta)\|^2\right)^2.
\end{align}
In the same manner, $F_3$ is further reduced to
\begin{align}
F_3=&\left(\mathbf{\dot{a}}^{H}\mathbf{V}_{A}\mathbf{R}_{yy}^{-1}\mathbf{V}_{A}^{H}\mathbf{a}\right)^2\\\nonumber
&=-\frac{4\pi^2\cos^2{\theta}}{M^2\lambda^2}\left(1-\frac{\Gamma}{\frac{M}{\gamma}+\Gamma}\right)^2\left(\mathbf{a}^{H}\mathbf{DBa}\right)^2.
\end{align}
Similar to the deriving process of $\mathbf{a}^H\mathbf{BDa}$ in (\ref{aHBDa}), we have
\begin{align}\label{aHDBa}
&\mathbf{a}^H\mathbf{DBa}\\\nonumber
&=(\mathbf{a}_{D}\otimes{\mathbf{a}_{A}})^H(\mathbf{I}_{K}\otimes{\mathbf{D}_A}+{\mathbf{D}_D}\otimes{\mathbf{I}_{M}})\\\nonumber
&\times(\mathbf{I}_{K}\otimes{\mathbf{E}_{M}})(\mathbf{a}_{D}\otimes{\mathbf{a}_{A}})\\\nonumber
&=\mathrm{Tr}(\mathbf{a}_D^H\mathbf{a}_D)\mathrm{Tr}(\mathbf{a}_A^H\mathbf{D}_A\mathbf{E}_M\mathbf{a}_A)+\mathrm{Tr}(\mathbf{a}_D^H\mathbf{D}_D\mathbf{a}_D)\\\nonumber
&\times\mathrm{Tr}(\mathbf{a}_A^H\mathbf{E}_M\mathbf{a}_A)\\\nonumber
&=Kg(\theta)\eta+\sum_{k=1}^{K}\mathbf{D}_{D,k}\|g(\theta)\|^2.
\end{align}
Placing (\ref{aHDBa}) in $F_3$ yields
\begin{align}\label{F-3-expres-1}
F_3&=\left(\mathbf{\dot{a}}^{H}\mathbf{V}_{A}\mathbf{R}_{yy}^{-1}\mathbf{V}_{A}^{H}\mathbf{a}\right)^{2}\\\nonumber
&=-\frac{4\pi^2\cos^2{\theta}}{M^2\lambda^2}\left(1-\frac{\Gamma}{\frac{M}{\gamma}+\Gamma}\right)^2\\\nonumber
&\times\left(Kg(\theta)\eta+\sum_{k=1}^{K}\mathbf{D}_{D,k}\|g(\theta)\|^2\right)^2
\end{align}
where
\begin{align}
\eta=\sum_{m=1}^{M}d_{m}e^{-j\frac{2\pi}{\lambda}d_m\sin{\theta}}.
\end{align}
Now, we go to calculate $F_2$. Making use of the identity (\ref{a-theta}), $F_2$ can be represented as follows
\begin{align}
F_2&=\left(\mathbf{a}^{H}\mathbf{V}_{A}\mathbf{R}_{yy}^{-1}\mathbf{V}_{A}^{H}\mathbf{a}\right)\left(\mathbf{\dot{a}}^{H}\mathbf{V}_{A}\mathbf{R}_{yy}^{-1}\mathbf{V}_{A}^{H}\mathbf{\dot{a}}\right)\\\nonumber
&=\frac{1}{M^2}(\Gamma-\frac{\Gamma^2}{\frac{M}{\gamma}+\Gamma})(\dot{\mathbf{a}}^H\mathbf{B}\dot{\mathbf{a}}-\frac{\dot{\mathbf{a}}^H\mathbf{Baa}^H\mathbf{B}\dot{\mathbf{a}}}{\frac{M}{\gamma}+\Gamma})\\\nonumber
&=\frac{4\pi^2\cos^2\theta}{\lambda^2M^2}\left(\Gamma-\frac{\Gamma^2}{\frac{M}{\gamma}+\Gamma}\right)\\\nonumber
&\times\left(\mathbf{a}^{H}\mathbf{D}^H\mathbf{BDa}-\frac{\mathbf{a}^H\mathbf{DBaa}^H\mathbf{BDa}}{\frac{M}{\gamma}+\Gamma}\right).
\end{align}
It is noted that, in order to compute $F_3$, we must derive $\mathbf{a}^{H}\mathbf{D}^H\mathbf{BDa}$ and $\mathbf{a}^H\mathbf{DBaa}^H\mathbf{BDa}$ firstly. Similar to the derivation of $\mathbf{a}^H\mathbf{BDa}$ and $\mathbf{a}^H\mathbf{DBa}$,
\begin{align}
&\mathbf{a}^H\mathbf{DBaa}^H\mathbf{BDa}\\\nonumber
&=(Kg(\theta)\eta+\sum_{k=1}^{K}\mathbf{D}_{D,k}\|g(\theta)\|^2)\\\nonumber
&\times({Kg(-\theta)\zeta+\sum_{k=1}^{K}\mathbf{D}_{D,k}\|g(\theta)\|^2})\\\nonumber
&=K^2\|g(\theta)\|^2\zeta\eta+K\|g(\theta)\|^2\sum_{k=1}^{K}\mathbf{D}_{D,k}(g(-\theta)\zeta+g(\theta)\eta)\\\nonumber
&+(\sum_{k=1}^{K}\mathbf{D}_{D,k}\|g(\theta)\|^2)^2,
\end{align}
and
\begin{align}
&\mathbf{a}^{H}\mathbf{D}^H\mathbf{BDa}\\\nonumber
&=(\mathbf{a}_D\otimes{\mathbf{a}_A})^H(\mathbf{I}_{K}\otimes{\mathbf{D}_A}+{\mathbf{D}_D}\otimes{\mathbf{I}_{M}})^H\\\nonumber
&\times(\mathbf{I}_{K}\otimes{\mathbf{E}_{M}})(\mathbf{I}_{K}\otimes{\mathbf{D}_A}+{\mathbf{D}_D}\otimes{\mathbf{I}_{M}})\\\nonumber
&\times(\mathbf{a}_D\otimes{\mathbf{a}_A})\\\nonumber
&=\mathrm{Tr}(\mathbf{a}_D\mathbf{a}_D^H)\mathrm{Tr}(\mathbf{a}_A^H\mathbf{D}_A^H\mathbf{E}_M\mathbf{D}_A\mathbf{a}_A)\\\nonumber
&+\mathrm{Tr}(\mathbf{a}_D^H\mathbf{D}_D^H\mathbf{a}_D)\mathrm{Tr}(\mathbf{a}_A^H\mathbf{E}_M\mathbf{D}_A\mathbf{a}_A)\\\nonumber
&+\mathrm{Tr}(\mathbf{a}_D^H\mathbf{D}_D\mathbf{a}_D)\mathrm{Tr}(\mathbf{a}_A^H\mathbf{D}_A^H\mathbf{E}_M\mathbf{a}_A)\\\nonumber
&+\mathrm{Tr}(\mathbf{a}_D^H\mathbf{D}_D^H\mathbf{D}_D\mathbf{a}_D)\mathrm{Tr}(\mathbf{a}_A^H\mathbf{E}_M\mathbf{a}_A)\\\nonumber
&=K\zeta\eta+\sum_{k=1}^{K}\mathbf{D}_{D,k}\left(g(\theta)\eta+g(-\theta)\zeta\right)\\\nonumber
&+\sum_{k=1}^{K}\mathbf{D}_{D,k}^2\|g(\theta)\|^2.
\end{align}
Considering $\zeta=\eta^H$, $F_2$ is written as
\begin{align}\label{F-2-expres-1}
F_2&=\left(\mathbf{a}^{H}\mathbf{V}_{A}\mathbf{R}_{yy}^{-1}\mathbf{V}_{A}^{H}\mathbf{a}\right)\left(\mathbf{\dot{a}}^{H}\mathbf{V}_{A}\mathbf{R}_{yy}^{-1}\mathbf{V}_{A}^{H}\mathbf{\dot{a}}\right)\\\nonumber
&=\frac{4\pi^2\cos^2\theta}{\lambda^2M^2}\left(\Gamma-\frac{\Gamma^2}{\frac{M}{\gamma}+\Gamma}\right)\\\nonumber
&\times\bigg(K\zeta\eta+2\sum_{k=1}^{K}\mathbf{D}_{D,k}{Re}\left\{g(\theta)\eta\right\}+\sum_{k=1}^{K}\mathbf{D}_{D,k}^2\|g(\theta)\|^2\\\nonumber
&-\big(K^2\|g(\theta)\|^2\|\eta\|^2+2K\|g(\theta)\|^2\sum_{k=1}^{K}\mathbf{D}_{D,k}{Re}\left\{g(\theta)\eta\right\}\\\nonumber
&+\|g(\theta)\|^4\sum_{k=1}^{K}(\mathbf{D}_{D,k})^2\big)\times(\frac{M}{\gamma}+\Gamma)^{-1}\bigg).
\end{align}
Finally, substituting $F_1$ in (\ref{F-1-expres-1}), $F_2$ in (\ref{F-2-expres-1}), and $F_3$ into (\ref{F-3-expres-1}) into (\ref{F-expre}) yields
\begin{align}\label{F-expre-1}
\mathbf{F}&=\frac{8\pi^2\cos^2{\theta}\gamma^2}{\lambda^2M^2}(1-\frac{\Gamma}{\frac{M}{\gamma}+\Gamma})(K\|g(\theta)\|^4\sum_{k=1}^{K}\mathbf{D}^2_{D,k}\\\nonumber
&-\|g(\theta)\|^4(\sum_{k=1}^{K}\mathbf{D}_{D,k})^2+K(1-\frac{\Gamma}{\frac{M}{\gamma}+\Gamma})\|g(\theta)\|^2\|\eta\|^2\\\nonumber
&+K^2(1-\frac{\Gamma}{\frac{M}{\gamma}+\Gamma}){Re}\left\{g^2(\theta)\eta\right\}).
\end{align}
According to $d$ in (\ref{d})
\begin{align}
d_m=(m-1)d,
\end{align}
and we derive that
\begin{align}
\Gamma=K\|g(\theta)\|^2.
\end{align}
Therefore, the FIM $\mathbf{F}$ in (\ref{F-expre-1}) is simplified to
\begin{align}\label{F-expre-2}
\mathbf{F}&=\frac{8\pi^2\cos^2{\theta}\gamma^2}{\lambda^2M(M+K\gamma\|g(\theta)\|^2)}\\\nonumber
&\times\Big(\frac{\|g(\theta)\|^4}{6}M^2K^2(K-1)(2K-1)d^2\\\nonumber
&-\frac{\|g(\theta)\|^4}{4}M^2K^2(K-1)^2d^2\\\nonumber
&+\frac{\|g(\theta)\|^2MK}{M+K\gamma\|g(\theta)\|^2}\|\eta\|^2\\\nonumber
&+\frac{MK^2}{M+K\gamma\|g(\theta)\|^2}{Re}\{g^2(\theta)\eta\}\Big).
\end{align}
The CRLB is given by
\begin{align}
CRLB=\frac{1}{L}\mathbf{F}^{-1}.
\end{align}
This completes our derivation of CRLB for hybrid structure. \hfill$\blacksquare$
\ifCLASSOPTIONcaptionsoff
  \newpage
\fi

\bibliographystyle{IEEEtran}
\bibliography{IEEEfull,cite}

\end{document}